\documentclass{moriond}
\usepackage{amsmath,amssymb}

\def\be{\begin{equation}}
\def\ee{\end{equation}}
\def\bea{\begin{eqnarray}}
\def\eea{\end{eqnarray}}

\def\gev{\,\text{GeV}}
\def\tev{\,\text{TeV}}
\def\B{\mathcal{B}}

\newcommand{\Photo}{}

\begin{document}
\vspace*{4cm}
\title{A MODEL TO ACCOMMODATE THE $B$-PHYSICS ANOMALIES}

\author{Damir Be\v{c}irevi\'c }

\address{Laboratoire de Physique Th\'eorique, CNRS et Univ.Paris-Sud, \\ 
Universit\'e Paris-Saclay, 91405 Orsay, France}

\maketitle\abstracts{
After briefly reviewing the status of the $B$-physics anomalies, describing the challenges of building a scenario of 
physics beyond the Standard Model that can accommodate the observed departures from lepton flavor universality, 
I discuss a new model, based on extending the Standard Model by two light [$\mathcal{O}(1\,\mathrm{TeV})$] 
scalar leptoquarks. That model, in addition to satisfying a number of flavor physics constraints both at low and high energies, also allows for a $SU(5)$ unification.}

\section{Introduction}

Over the past several years we witnessed a growing experimental evidence of the lepton flavor universality violation (LFUV) in the decays of $B$-mesons. 
In particular, the measured ratios~\footnote{$\B$ and $\B'$ stand for the full and partial branching fraction, respectively.}
\bea
&&R_{D^{(\ast )}}= \left. \frac{\B (B\to D^{(\ast )} \tau \nu_\tau)}{\B (B\to D^{(\ast )} l \nu_l)}\right|_{l\in {e,\mu}},\cr
&& R_{K^{(\ast )}}=   \frac{\B' (B\to K^{(\ast )} \mu \mu)}{\B' (B\to K^{(\ast )} ee)}, 
\eea 
appear to be different from the values predicted in the Standard Model (SM). While for the SM tree-level decays $b\to c\ell\bar \nu_\ell$ the observed ratios are larger than predicted ($R_{D^{(\ast )}}^\mathrm{exp}>R_{D^{(\ast )}}^\mathrm{SM}$) to $\approx 3\sigma$ level, similar ratios for the $b\to s\ell\ell$ decays, induced by quantum loops in the SM, turn out to be smaller than predictions ($R_{K^{(\ast )}}^\mathrm{exp}<R_{K^{(\ast )}}^\mathrm{SM}$) to $\approx 2.5\sigma$. These deviations, often referred to as the {\it $B$-physics anomalies}, need further experimental verification and improvements that are supposed to be ensured in the new runs at LHC, and especially at the $B$-factory at KEK, Belle~II. If $R_{D^{(\ast )}}^\mathrm{exp}>R_{D^{(\ast )}}^\mathrm{SM}$ and/or $R_{K^{(\ast )}}^\mathrm{exp}<R_{K^{(\ast )}}^\mathrm{SM}$ remain true to at least $5\,\sigma$, that would constitute a clear signal of physics beyond the SM. Building a model of new physics (NP) that can describe both types of $B$-physics anomalies turns out to be a very challenging task. In this write-up we 
will elaborate on a new model which -- to date -- arguably offers the most appealing solution to the problem discussed above, cf. Ref.~\cite{BDFFKS}.

\subsection{Remarks on the SM uncertainties}
The SM estimates of the above-mentioned LFUV ratios are rather robust even though a full theoretical control over the hadronic uncertainties is still lacking. 
One of the reasons to prefer working with ratios is that many hadronic uncertainties actually cancel.
The residual errors, arising mostly from the hadronic form factors, are however still significant and can be reduced model-independently through numerical simulations 
of QCD on the lattice. 
With regards to that statement, we first remark that the form factors $ f_{+,0}(q^2)$, relevant to the $B\to D \ell \nu_\ell$ decays, have been computed on the lattice by two different 
lattice collaborations~\cite{Lattice:2015rga} in the region of $q^2$'s close to $q^2_\mathrm{max}=(m_B-m_D)^2$.~\footnote{Hadronic matrix elements of the weak charged current is expressed in terms of two and four form factors, depending on the spin of the final state, $\langle D|\bar c\gamma^\mu_L b|B\rangle \propto f_+(q^2), f_0(q^2)$, and $\langle D^\ast |\bar c\gamma^\mu_L b|B\rangle \propto A_1(q^2),A_2(q^2),A_0(q^2),V(q^2)$. In the SM the scalar and pseudoscalar form factors, $f_0(q^2)$ and $A_0(q^2)$, are only relevant if $\ell =\tau$.} Extrapolation to $q^2_\mathrm{min}=m_\ell^2$ is highly constrained by the condition that the two form factors are equal to each other at $q^2=0$, $f_+(0)=f_0(0)$. 
The situation is much less convenient in the case of $B\to D^\ast \ell \nu_\ell$ for which the estimate of the form factors away from $q^2_\mathrm{max}=(m_B-m_{D^\ast})^2$ has been made only in the quenched approximation~\cite{deDivitiis:2008df}. In such a situation one has to make two assumptions: (i) The shape of the form factor $A_1(q^2)$, which in the heavy quark limit corresponds to the celebrated Isgur-Wise function, is truncated to a quadratic function of $w-1$, where the relative velocity is related to $q^2$ via $w= (m_B^2+m_{D^\ast}^2-q^2)/( 2 m_B m_{D^\ast})$. The unitarity bound 
on the decay amplitude can be used to model the shape of $A_1(w)$ as to eliminate one parameter, cf. Refs.~\cite{CLN}, so that only one parameter is needed to describe both the linear and the quadratic dependence in $(w-1)$. (ii) The ratios between other form factors and $A_1(q^2)$ are also parametrized as quadratic functions of $w-1$. Those shapes, in the SM, can be extracted from the experimentally measured angular distribution of $B\to D^\ast (\to D\pi) l\nu_l$ decay ($l\in {e,\mu}$), which has been done in Refs.~\cite{EXP}. In other words, as of now, we need to assume that the departure from lepton flavor universality in $R_{D^\ast}$ is (almost) entirely due to $\B (B\to D^{\ast } \tau \nu_\tau)^\mathrm{exp} > \B (B\to D^{\ast } \tau \nu_\tau)^\mathrm{SM}$. 
Notice also that the (pseudo-)scalar form factor $A_0(q^2)$ has never been computed on the lattice. Its value has only been deduced from considerations based on heavy quark effective theory (HQET).  For a recent discussion regarding the form factor values and their shapes see Ref.~\cite{Bigi:2017jbd}.

As for $B\to K^{(\ast )} l^+l^-$ decays the corresponding form factors have not been computed on the lattice for small $q^2\in (1,6)\gev^2$, i.e. in the region in which the ratios $R_{K^{(\ast )}}$ have been measured. Instead, they have been computed for large $q^2$-values and then extrapolated to the low $q^2$'s~\cite{Bouchard:2013pna}. Due to the fact that the form factors at low $q^2$'s are flat functions of $q^2$, it is extremely important to compute their values at $q^2\approx 0$. Small deviations from flat behavior can be computed in any given model. A model which captures most features of QCD and allows to compute the form factor values at low $q^2$'s is the so called QCD sum rules near the light cone (LCSR). The size of systematic uncertainties of the results of LCSR is often subject of controversies. A common practice is to use LCSR result and not question the robustness of the estimated uncertainties.

Finally, and before closing this set of remarks, we need to emphasize that an important source of uncertainty come from the soft photon radiation, which in the case of $R_{K^{(\ast )}}$ is actually the main source of theory error~\cite{Bordone:2016gaq}. 
Early estimates of that source of uncertainties to $R_{D^{(\ast )}}$ have been discussed in Ref.~\cite{Becirevic:2009fy}.

\subsection{$B$-physics anomalies in numbers}

We close this section by quoting the experimental values: 
\bea\label{R:exp}
&&R_D= 0.41(5), \qquad R_{D^\ast}= 0.30(2), \qquad R_{J/\psi}=0.71(25),\nonumber \\
&&R_K^{[1,6]}=0.74(9),\qquad 
R_{K^\ast}^{[1.1,6]}=0.71(10) ,\qquad  
R_{K^\ast}^{[0.045,1.1]}=0.68(9),
\eea
$R_{D}$ and $R_{D^{\ast }}$ have been measured by several collaborations and the above results are the average values~\cite{Amhis:2016xyh}. 
The value of $R_{J/\psi}$ has been taken from Ref.~\cite{Aaij:2017tyk}.
$R_K$ and $R_{K^{\ast }}$, at this level of precision, have only been measured by LHCb the results of which were reported in Refs.~\cite{ Aaij:2014ora,Aaij:2017vbb}, respectively.
Notice also that the superscript in $R_{K^{(\ast )}}$ indicate the range of $q^2$'s over which the partial branching fractions have been measured.

The above values are to be compared with those predicted in the SM, namely,
\bea\label{R:theo}
&&R_D= 0.30(1), \qquad R_{D^\ast}= 0.26(1), \qquad R_{J/\psi}=0.23(1),\nonumber \\
&&R_K^{[1,6]}=1.00(1),\qquad 
R_{K^\ast}^{[1.1,6]}=1.00(1) ,\qquad  
R_{K^\ast}^{[0.045,1.1]}=0.66(3).
\eea
Hadronic uncertainties in $R_D$ have been tamed by lattice QCD in Ref.~\cite{Lattice:2015rga}. The value of $R_{D^\star}$ is taken from Ref.~\cite{Bigi:2017jbd} in which the previous estimate~\cite{Fajfer:2012vx} has been updated. $R_{J/\psi}$ was estimated by combining the lattice QCD and QCD sum rules~\cite{us}. The SM values for $R_{K^{(\ast )}}$ were taken from Ref.~\cite{Bordone:2016gaq}.

The difference between the numbers shown in Eq.~\eqref{R:exp} and those in Eq.~\eqref{R:theo} is what is known today as the $B$-physics anomalies. Other similar channels are nowadays being explored experimentally and we can soon expect the results for $R_{D_s}$, $R_{\Lambda_c^{(\ast )}}$, $R_\phi$ and others~\cite{langenbruch}.  

\subsection{What New Physics Scenario for $B$-physics anomalies?}

Clearly, LFUV cannot be accommodated in the SM and one has to look for a scenario beyond the SM. Since this part is abundantly discussed 
 in several talks at this conference, the reader is referred to Refs.~\cite{LFUV-moriond}. I would only mention that building a model to accommodate 
 the above anomalies is extremely challenging because of the numerous constraints arising (i) from the direct searches in colliders (LHC, in particular), (ii) from 
 the low-energy physics observables, and (iii) from perturbative unitarity. 
Models that can accommodate the anomalies in the processes governed by the flavor changing charged currents ($b\to c\ell\bar\nu_\ell$) usually fail in describing the anomalies observed in the processes governed by the flavor changing neutral currents ($b\to s\ell^+\ell^-$) and vice versa. 
One class of models which seems to be appropriate for describing the LFUV effects are those which involve one or more light leptoquark (LQ) state(s), where by {\it light} a mass $m_{LQ} = \mathcal{O}(1\tev )$ is assumed.~\footnote{The main reason for insisting on $m_{LQ} = \mathcal{O}(1\tev )$ is that its presence can be directly checked on at the LHC.} It appears, however, that no model with a single light scalar LQ can provide a solution to both kinds of $B$-physics anomalies. One way out is to build a scenario with a \underline{light} vector LQ, but the problem one encounters in those models is that a new theory is not renormalizable and the UV-completion should be devised. Consequently a model with light vector leptoquark(s) require more particles and parameters resulting in a not so elegant a solution to the problem in hands. Since we want a model to be minimalistic (in terms of new parameters), the most favorable  situations in that respect are those in which we combine two light scalar LQ's. 

\subsection{New Viable Model}

Our model at the $\tev$-scale -- in flavor basis -- can be described by 
\bea
\mathcal{L} \ {\supset} \  {\color{red} y_{R}^{ij} }\, \bar{Q}_i {\color{blue}R_2} \ell_{Rj}  +{\color{red} y_L^{ij} }\, \bar{u}_{Ri} {\color{blue} \widetilde{R}_2^\dagger } L_j + {\color{red} y^{ij} } \, \bar{Q}^C_{i} i \tau_2 ( \tau_k {\color{blue}S^k_3}) L_{j}+\mathrm{h.c.}\,,
\eea
where besides the quark and lepton fields in usual notation we also introduced the so called $R_2$ and $S_3$ scalar LQ's, 
which carry the SM quantum numbers $R_2=(3,2,7/6)$ and $S_3=(\bar 3,3,1/3)$, referring to $SU(3)_c$, $SU(2)_L$ 
and $U(1)_Y$, respectively. $y$ and $y_{L,R}$ are the matrices of Yukawa couplings. Specifying their content means specifying the model. Before doing that, we first 
rotate the above Lagrangian to the fermion mass eigenstate basis and get
\begin{align}\label{model}
\begin{split}
\mathcal{L}\ \supset \ &{\color{red}  (V_\mathrm{CKM} \, y_R \, E_R^\dagger)^{ij}}\,\bar{u}^\prime_{Li}\ell^\prime_{Rj} {\color{blue}  R_2^{(5/3)} }+ {\color{red} (y_R \, E_R^\dagger)^{ij}}\, \bar{d}^\prime_{Li}\ell^\prime_{Rj} {\color{blue} R_2^{(2/3)} }\\
&+{\color{red} (U_R \, y_L \, U_\mathrm{PMNS})^{ij}}\,\bar{u}^\prime_{Ri} \nu^\prime_{Lj}  {\color{blue} R_2^{(2/3)} }-{\color{red}  (U_R \, y_L)^{ij} }\, \bar{u}^\prime_{Ri}\ell^\prime_{Lj}  {\color{blue} R_2^{(5/3)} }\\
&-{\color{red} (y \,U_\mathrm{PMNS})^{ij}}\, \bar{d}^{\prime C}_{Li} \nu^\prime_{Lj}  {\color{blue} S_3^{(1/3)}} - \sqrt{2}{\color{red}  \,y^{ij} }\,\bar{d}^{\prime C}_{Li} \ell^\prime_{Lj}  {\color{blue} S_3^{(4/3)}}\\
&\hspace*{-5mm}+\sqrt{2}{\color{red} (V_\mathrm{CKM}^* \, y \,U_\mathrm{PMNS})_{ij}}\, \bar{u}^{\prime C}_{Li} \nu^\prime_{Lj}  {\color{blue} S_3^{(-2/3)}}-{\color{red} (V_\mathrm{CKM}^* \, y)_{ij} }\,\bar{u}^{\prime C}_{Li} \ell^\prime_{Lj}  {\color{blue} S_3^{(1/3)}} + \mathrm{h.c.}\,,
\end{split} 
\end{align}
where we recognize the usual quark and lepton mixing matrices, $V_\mathrm{CKM}$ and $U_\mathrm{PMNS}$. Note that the superscripts of the LQ field correspond to their electric charge. 
Our (minimalistic) choice of Yukawa matrices that can accommodate the $B$-physics anomalies is:
\begin{equation}
\label{eq:yL-yR}
y_R \, E_R^\dagger = \begin{pmatrix}
0 & 0 & 0\\ 
0 & 0 & 0\\ 
0 & 0 & y_R^{b\tau}
\end{pmatrix} \,,~
U_R \, y_L = \begin{pmatrix}
0 & 0 & 0\\ 
0 & y_L^{c\mu} & y_L^{c\tau}\\ 
0 & 0 & 0
\end{pmatrix}\,,~
U_R = \begin{pmatrix}
1 & 0 & 0\\ 
0 & \cos \theta & -\sin \theta\\ 
0 & \sin \theta & \cos \theta
\end{pmatrix}\,,
\end{equation}
satisfying $y_R=y_R^T$ and $y=-y_L$. The model thus involves six new parameters: $m_{R_2}$, $m_{S_3}$, $y_R^{b\tau}$, $y_L^{c\mu}$, $y_L^{c\tau}$, and $\theta$. 
From the analysis of compatibility with various experimental data we find that $\theta \approx \pi/2$ and that $y_R^{b\tau}$ is mostly imaginary. 

\subsection{$R_K$ and $R_{K^\ast}$}

A common way to describe the $b\to s\ell\ell$ decays is to devise a low energy effective theory with   
\begin{align}
&\mathcal{H}_\mathrm{eff}\supset -\frac{4 G_F}{\sqrt{2}}V_{tb}V_{ts}^\ast\,  \Big{(}C_9(\mu)\mathcal{O}_9+C_{10}(\mu)\mathcal{O}_{10} \Big{)} + \mathrm{h.c.}\,,\nonumber\\
{\mathcal{O}_{9}} = {e^2\over (4\pi)^2}&{(\bar{s}\gamma_\mu P_{L} b)(\bar{\ell}\gamma^\mu\ell)},\qquad \qquad 
 {\mathcal{O}_{10}} = {e^2\over (4\pi)^2} {(\bar{s}\gamma_\mu P_{L} b)(\bar{\ell}\gamma^\mu\gamma^5\ell)} ,
 \end{align}
where we display only the contributions which are of interest to our considerations here. The Wilson coefficients $C_{9,10}$ are obtained by matching the effective and full theories in the SM. To those one adds the contributions arising from our model~\eqref{model} for which we obtain, 
\bea
C_9^\mathrm{\mu\mu}=-C_{10}^\mathrm{\mu\mu}  = \dfrac{\pi v^2}{ V_{tb}V_{ts}^\ast \alpha_{\mathrm{em}}} \dfrac{y_{b\mu} y_{s\mu}^\ast}{m_{S_3}^2}\ .
\eea
We see that the LFUV is assumed to arise from the NP couplings to muons and not to electrons and only $S_3$ can modify $b\to s\mu\mu$. This is consistent with observations based on the results of global fits to many $b\to s\ell\ell$ observables, cf. Fig.~9 in Ref.~\cite{Descotes-Genon:2015uva}. After comparing theory with experiment for $\B (B_s\to \mu\mu)$, $R_K$ and $R_{K^\ast}$ we get $C_9^\mathrm{\mu\mu}\in (-0.85,-0.50)$, where in the theory estimate of $R_K$ and $R_{K^\ast}$ we assume the validity of the values and uncertainties of the form factors as estimated by means of LCSR~\cite{Ball:2004rg}.

\subsection{$R_D$ and $R_{D^\ast}$}
\begin{figure}
\begin{minipage}{0.5\linewidth}
\centerline{\includegraphics[width=0.9\linewidth]{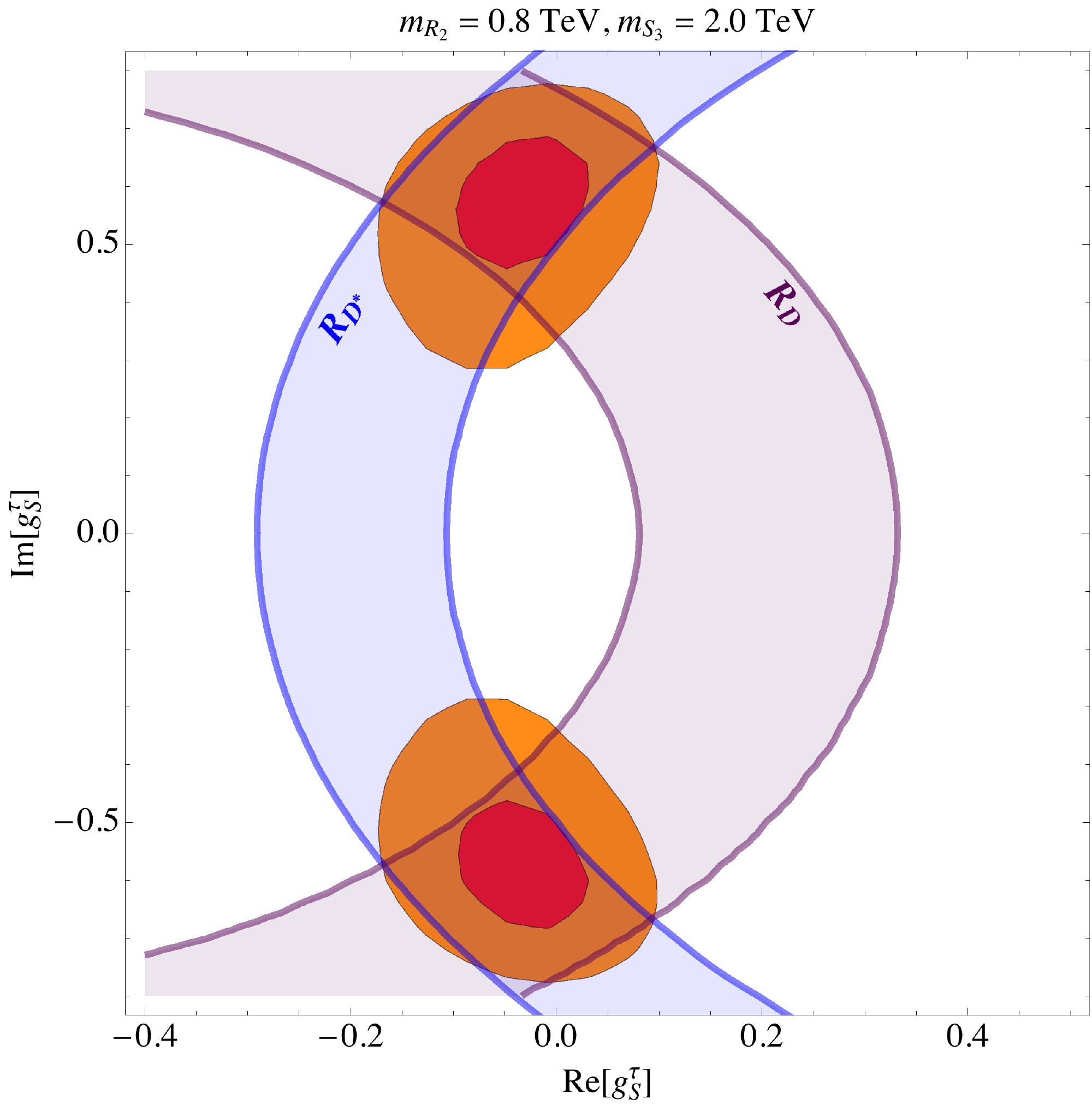}}
\end{minipage}
\hfill
\begin{minipage}{0.5\linewidth}
\centerline{\includegraphics[width=0.98\linewidth]{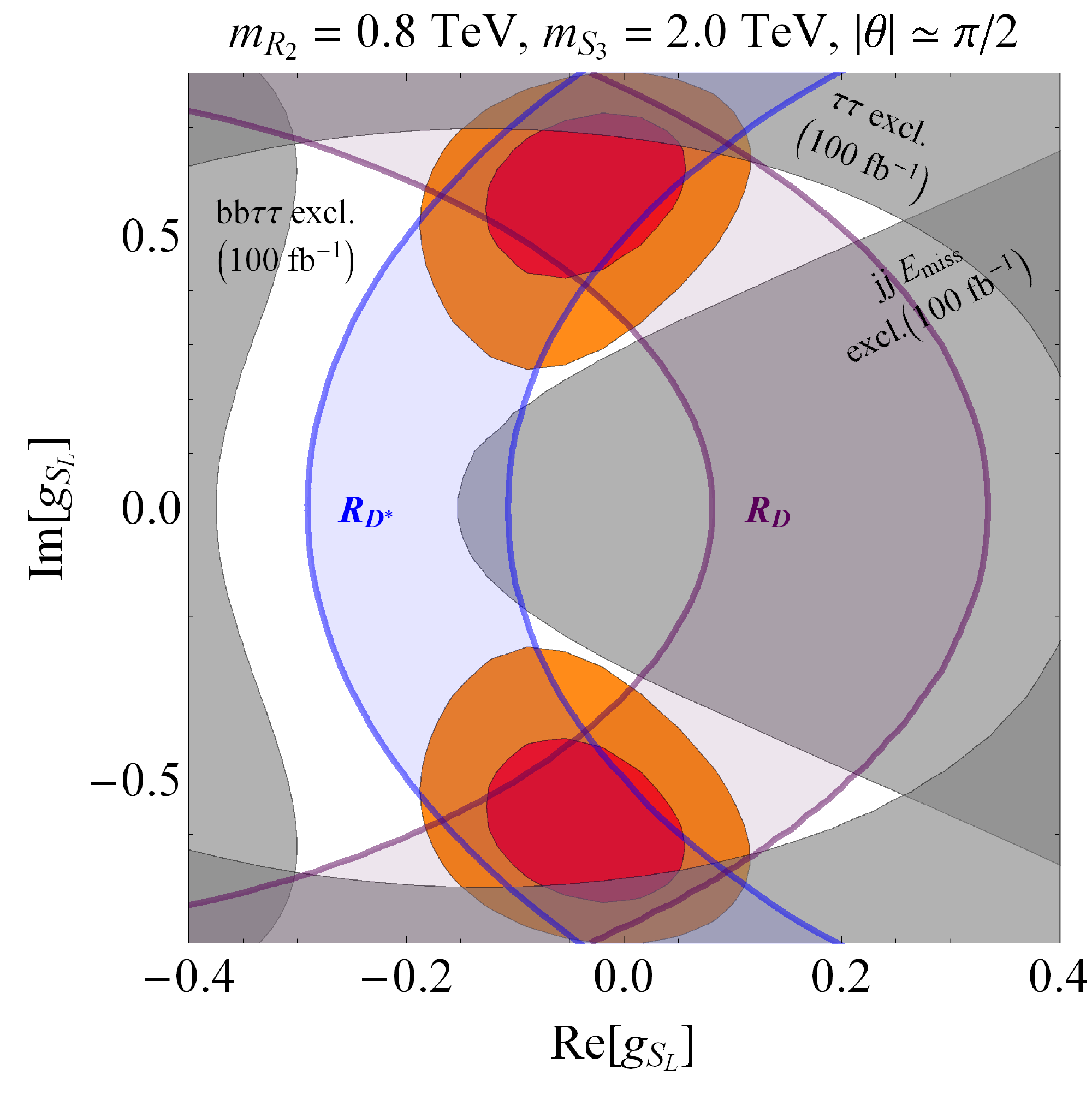}}
\end{minipage}
\caption[]{Constraints from the low energy data on $\mathrm{Re}[ g_S]$ and $\mathrm{Im}[ g_S]$, fully compatible with $R_D$ and $R_{D^\ast}$ which are also shown in the plot. In the right plot the same region of selected points is combined with projected constraints arising from direct searches. }
\label{fig:1}
\end{figure}

We proceed as before and build a low energy effective theory to describe $b\to c\ell\nu_\ell$ via,
 \begin{align}
\begin{split}
{\cal L}_{\mathrm{eff}} \supset -\frac{4 \, G_F}{\sqrt{2}} V_{cb} \big{[} (1+{g_V}) 
(\bar{u}_L \gamma_\mu d_L)&(\bar{\ell}_L \gamma^\mu \nu_{L}) 
+  g_S(\mu)\, (\bar{u}_R  d_L)(\bar{\ell}_R \nu_L) \\[0.01em]
&+ {g_T(\mu)}\, (\bar{u}_R  \sigma_{\mu \nu} d_L) (\bar{\ell}_R \sigma^{\mu \nu}\nu_L) 
\big{]}  + \mathrm{h.c.}
\label{eq:hamiltonian-semilep}
\end{split}
\end{align}
After a quick inspection of Eq.~\eqref{model} we can easily identify the NP couplings $g_{V,S,T}$ with 
\bea\label{R22}
\left. g_S(\mu) = 4 \,g_T(\mu) = \frac{y_L^{u\ell^\prime}\, \left(y_R^{d\ell}\right)^{\ast}}{4 \sqrt{2} \, m_{R_2}^2 \, G_F V_{ud}} \, \right|_{\mu = m_{R_2}} ,\qquad 
g_V = -\dfrac{y_{d\ell^\prime} \left(V y^\ast\right)_{u\ell}}{4\sqrt{2} \, m_{S_3}^2 G_F V_{ud}},
\eea
with $u \in \{u,c\}$, $d \in \{s,b\}$,  $\ell^{(\prime)} \in \{\mu,\tau\}$. $g_V$ appears to be tiny and can be safely neglected, while the LFUV comes mostly from a non-zero value of $g_S$, which in turn arises from couplings to $R_2$. 
A simultaneous fit with experimental data for $R_D$ and $R_{D^\ast}$ suggests that $g_S$ should be complex, cf. Fig.~\ref{fig:1}. In obtaining the plausible region of $g_S\neq 0$ we added many more observables in the global fit, such as $\Delta m_{B_s}$, $\Gamma(Z\to \ell\ell)$, $\Gamma(Z\to \nu\nu)$, as well as the LEP bound on $\mathcal{B}(\tau\to  \mu \phi)$, and the experimental values for ${\Gamma(K^-\to e^- \bar{\nu})/\Gamma(K^-\to \mu^- \bar{\nu})}$ and ${\Gamma(B\to D^{(\ast ) } \mu \bar{\nu})/\Gamma(B\to D^{(\ast ) } e \bar{\nu})}$. We would like to stress the importance of the experimental bound on $\mathcal{B}(\tau\to  \mu \phi)$ which helps selecting the larger of the two plausible solutions, $\theta \approx 0$ and $\theta \approx \pi/2$. 

Notice also that the electroweak corrections mix $g_S$ and $g_T$ as discussed in Ref.~\cite{Gonzalez-Alonso:2017iyc}. We took that effect properly into account -- in addition to QCD -- while running between $\mu=m_b \leftrightarrow \mu=m_{R_2}$. 
By setting e.g. $\mathrm{Re}[g_S]=0$, we obtain $\mathrm{Im}[ g_S]= 0.59^{+0.13}_{-0.14}$, to $1\sigma$, which is attributed to $y_R^{b\tau}$. We checked that this NP phase does not entail any observable CP-violating effect. We also checked that our model is fully consistent with the measured $D^0-\overline D^0$ mixing parameters.

\subsection{Perturbativity}

In addition to all of the above, we also require all of our non-zero Yukawa couplings to be smaller than $\sqrt{4\pi}$. While this is clearly realized from our fits with low energy data, venturing to a scenario of grand unification (GUT) might create problems for some Yukawa couplings which can become far too large when running from $\Lambda \sim 1\tev$ to $\Lambda_\mathrm{GUT}\simeq 5\times 10^{15}\gev$~\cite{Wise:2014oea}. 

Furthermore, it is important to have small or moderate values of $y_R^{b\tau}$ and $y_L^{c\tau}$ because they 
could otherwise generate important modifications to the running of the third generation of the SM fermions.

\subsection{Constraints from direct searches at the LHC}
\begin{figure}
\begin{minipage}{0.5\linewidth}
\centerline{\includegraphics[width=.93\linewidth]{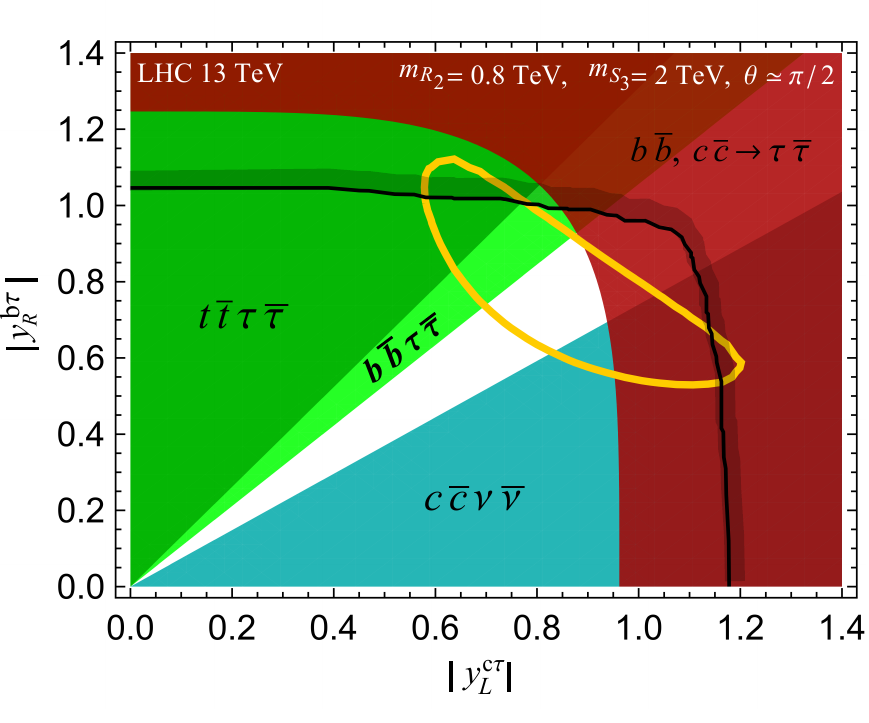}}
\end{minipage}
\hfill
\begin{minipage}{0.5\linewidth}
\centerline{\includegraphics[width=0.8\linewidth]{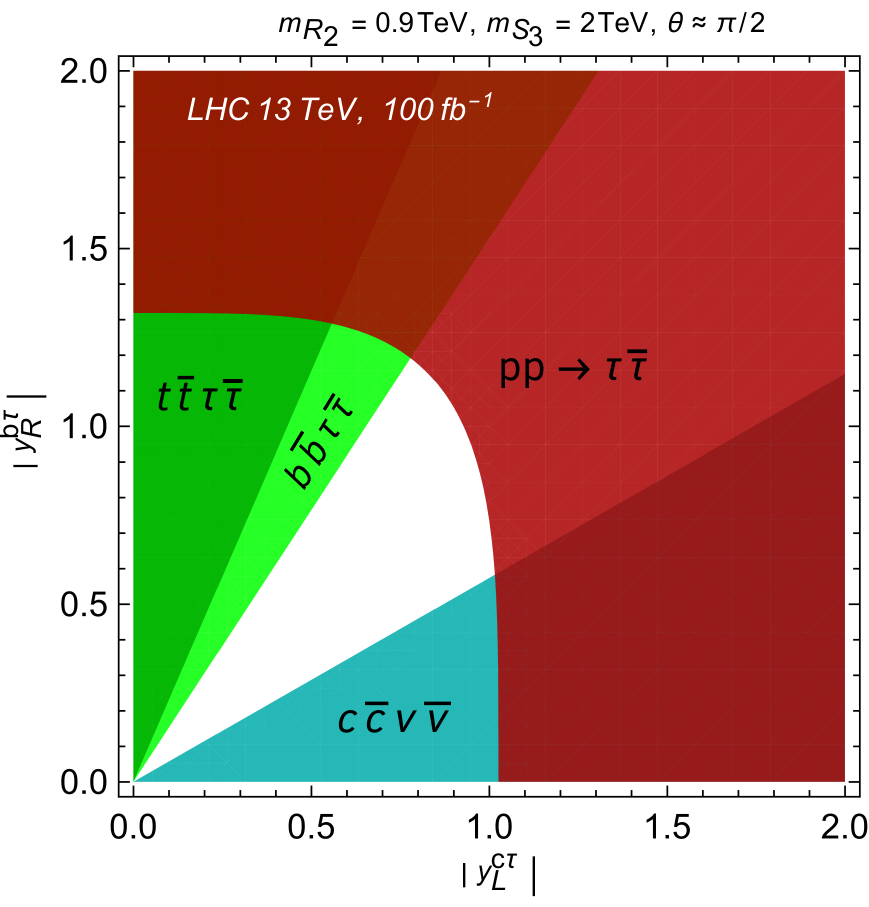}}
\end{minipage}
\caption[]{In the left (right) plot we show the constraints on the Yukawa couplings coming from the direct searches using the data projected to $100\ \mathrm{fb}^{-1}$ and by fixing $m_{R_2}$ to $800$ ($900$)~GeV. In the left plot we also show the curve corresponding to the perturbative unitarity, as well as the ``ellipse-like" shape which comes from the fit with low energy flavor data to $1\,\sigma$.}
\label{fig:2}
\end{figure}

As it was shown in Ref.~\cite{Faroughy:2016osc}, important constraints on the Yukawa couplings can be inferred from direct searches in colliders. 
Couplings to a light LQ can, for example, result in a significant deviation of the tails of the $p_T$ distribution of $\sigma(pp\to \ell\ell)$. 
Since in our model the Yukawa couplings are small, the current data do not provide constraints competitive with those we discussed before.  
However, after projecting to $100~\mathrm{fb}^{-1}$ from the tails of 
$pp\to \tau\tau$ one can extract efficient bounds on Yukawa couplings, which in our case ($\theta \approx \pi/2$) are important for $R_2$ and only 
marginal for $S_3$. We checked that the $pp\to \mu\mu$ data are not very useful to constrain our model. Instead, the experimental bounds on the LQ pair productions 
can bring very significant constraints, especially if focusing onto the final states $t\bar t\tau\bar \tau$,  $b\bar b\tau\bar \tau$,  $c\bar c\nu\bar \nu$. Relevant constraints are shown in Fig.~\ref{fig:2} (left plot) where we also show the perturbativity bound, as well as the bounds from the low energy flavor observables. 

\subsection{Predictions}

\begin{figure}
\begin{minipage}{0.5\linewidth}
\centerline{\includegraphics[width=.85\linewidth]{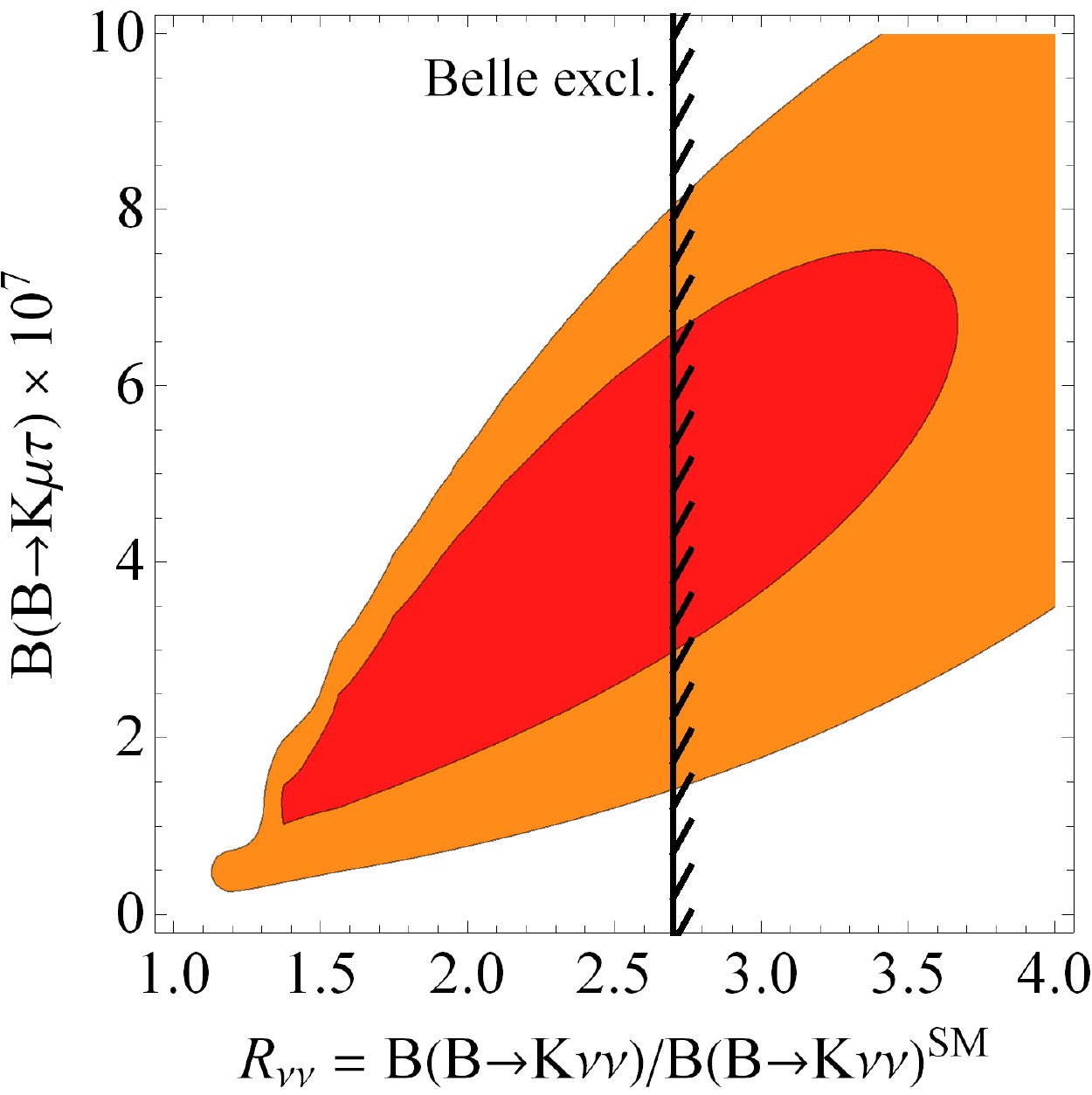}}
\end{minipage}
\hfill
\begin{minipage}{0.5\linewidth}
\centerline{\includegraphics[width=0.85\linewidth]{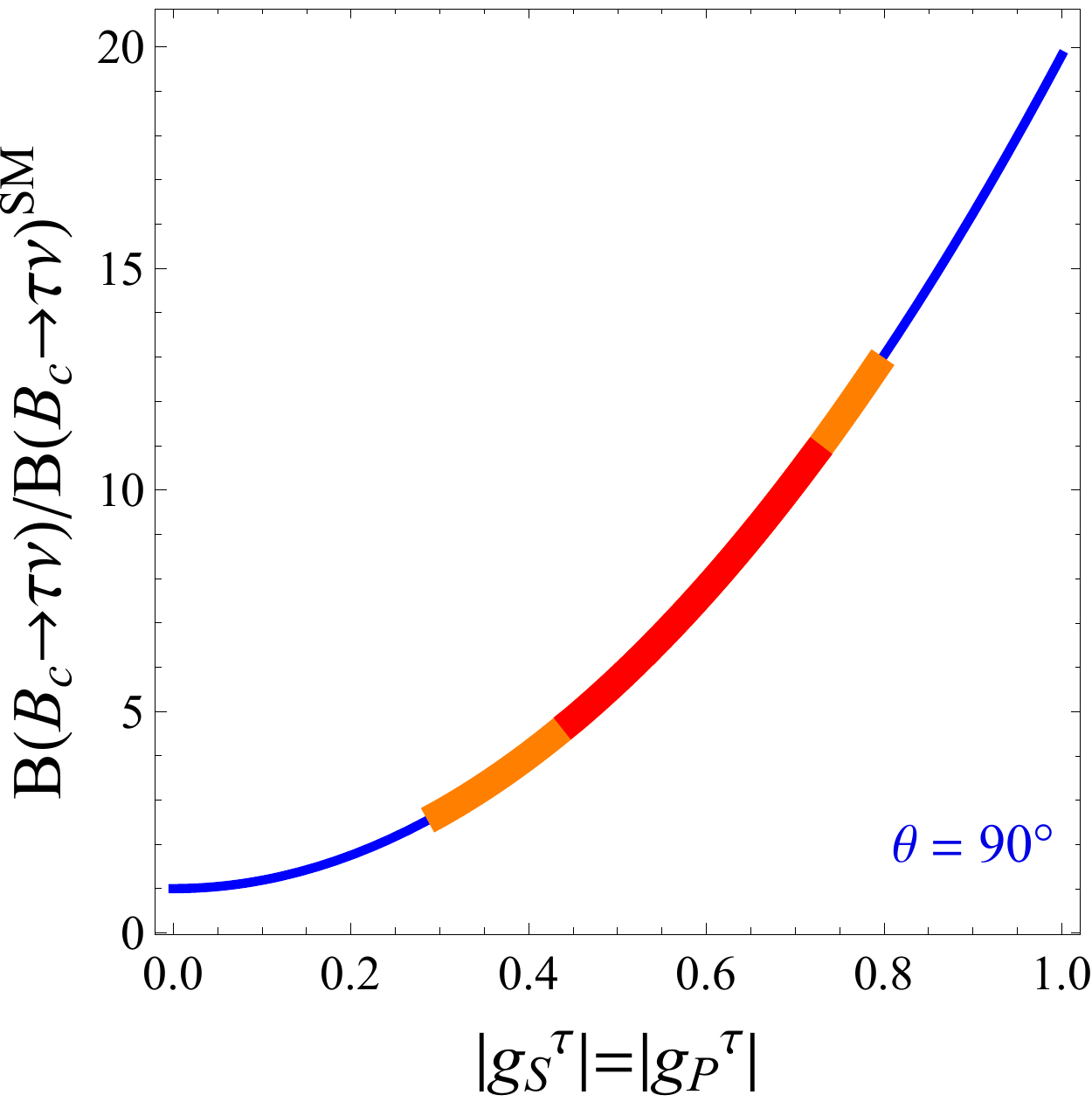}}
\end{minipage}
\caption[]{Predictions based on the model proposed here: Left panel shows the correlation between $\B (B\to K\mu\tau)$ and $R_{\nu\nu}$ which allows one to deduce the upper and lower bound on $\B (B\to K\mu\tau)$, as discussed in the text. Right panel shows the enhancement of $\B (B_c\to \tau\bar \nu_\tau )$ with respect to its SM value (red and orange regions correspond to $1$ and $2\,\sigma$ constraint, respectively). }
\label{fig:3}
\end{figure}

We end this Section by spelling out several predictions of our model. 
As discussed in our previous papers, the possibility of lepton flavor violation (LFV) is very appealing since these decay modes are likely to be studied at LHCb and at Belle~II. 
We notice an important correlation with $R_{\nu\nu}^{(\ast )} = \B (B\to K^{(\ast )}\nu\nu)/\B (B\to K^{(\ast )}\nu\nu)^\mathrm{SM}$ which has been recently bounded experimentally~\cite{Grygier:2017tzo}, so that 
we get both the upper and lower bound on the LFV modes, cf. Fig.~\ref{fig:3}, i.e. 
\bea
1.1 \times 10^{-7} \lesssim \B (B\to K\mu\tau)\lesssim  6.5 \times 10^{-7}\,,
\eea
however, two orders of magnitude smaller than the current experimental bound. Note that the other similar LFV modes 
are related to the above one via $\B (B\to K^\ast\mu\tau)\approx 1.9 \times  \B (B\to K\mu\tau)$, and $\B (B_s\to \mu\tau)\approx 0.9 \times  \B (B\to K\mu\tau)$, see Ref.~\cite{Becirevic:2016zri}. We also get interesting lower ($1~\sigma$) bounds  $\B (B\to K^\ast\mu\tau)\gtrsim 1.5 \times 10^{-8}$, and $\B (B\to K^\ast\nu \nu)\gtrsim 1.3 \times \B (B\to K^{(\ast )}\nu\nu$.

We should also note that $\B (B_c\to \tau\bar \nu_\tau )/\B (B_c\to \tau\bar \nu_\tau )^\mathrm{SM}$ gets enhanced by a factor between $5$ and $11$, within bounds discussed in Ref.~\cite{Alonso:2016oyd}.

\subsection{$SU(5)$ embedding: GUT -- The Gravy }

One might wonder about our motivation for choosing the Yukawa matrices as in Eq.~\eqref{eq:yL-yR}, where we also required $y_R=y_R^T$ and $y=-y_L$. 
That choice was actually driven by the possibility of embedding our model in a plausible $SU(5)$ GUT scenario. 
We should note that in $SU(5)$ the SM matter fields are in $\mathbf{5}$ and  $\mathbf{10}$ multiplets, while the scalar leptoquarks live in 
$R_2\in \mathbf{45},\mathbf{50}$ and $S_3\in \mathbf{45}$. In order to comply with the bounds on the proton lifetime, the contractions 
  $z\cdot \mathbf{10}_i\mathbf{10}_j \mathbf{45}$ are constrained to extremely small values ($z\ll 1$), which we can safely set to zero ($z = 0$)~\cite{Dorsner:2017ufx}. Other available operators are
\begin{align}
\mathbf{10}_i\mathbf{5}_j  \mathbf{45} & : \quad  y^{RL}_{2\ ij} \ \overline u^i_R  R_2^a \varepsilon^{ab} L_L^{j,b}, \quad 
 y^{LL}_{3 ij} \  {\overline{Q^c}}^{i,a}_L \varepsilon^{ab} (\tau^k  {S_3^k})^{bc} L_L^{j,c}, \nonumber\\
\mathbf{10}_i\mathbf{10}_j \mathbf{50}& : \quad  y^{LR}_{2\ ij}\  \overline e^i_R  R_2^{a\, \ast} Q_L^{j,a},
\end{align}
 where $i,j$ are the generation indices. While breaking $SU(5)$ down to SM, the two $R_2$ mix in such a way that one can be light and the other one very heavy. 
 In that way we remain with two light LQ's, one being $R_2$ and the other $S_3$. This explains our motivation for Eqs.~\eqref{model} and \eqref{eq:yL-yR}.

\section{Conclusion}

\begin{itemize}
\item Currently there is a considerable effort to build a viable NP model that would accommodate the $B$-physics anomalies, but remains consistent with all other measured flavor observables. That task is highly nontrivial as we are in the era in which the model building efforts are data driven.

\item We propose a minimalistic model which involves two scalar leptoquarks of mass $\mathcal{O}(1\tev)$ and which complies with all of the available low energy flavor physics constraints. A peculiarity is that one of the Yukawa couplings is complex. The corresponding NP phase does not result in any visible CP-violating effects. Furthermore, regarding the NP contribution to $b\to c\tau \nu$ modes, our model is not of the ``V-A"-type. Instead in generates terms proportional to (pseudo-)scalar and to tensor operators in a peculiar combination, cf. Eq.~\eqref{R22}.

\item The high-$p_T$ tail of $pp\to \ell\ell$ can be sensitive to the presence of light LQ's. Together with the bounds concerning the leptoquark pair production, we show that our model is expected to stay consistent with direct searches after projecting to $100\,\mathrm{fb}^{-1}$ of data. Since such data are becoming increasingly important, it is essential to correctly estimate the experimental background before promoting the experimental bounds to the stringent constraints on the Yukawa couplings. 

\item We also show that the model we build is inspired by the $SU(5)$ unification. By relying on the 1-loop running, we checked that our Yukawa couplings satisfy the perturbativity all the way up to the unification scale which we set at $\Lambda_\mathrm{GUT}\simeq 5\times 10^{15}\gev$.
  
\end{itemize}

\section*{Acknowledgments}
{
I would like to thank the organizers for the invitation and especially to Svjetlana Fajfer, Ilja Dor\v{s}ner, Darius Faroughy, Nejc Ko\v{s}nik and Olcyr Sumensari from whom and with whom I learned so much about the subject of this talk.}


\section*{References}

\begin{thebibliography}{99}

\bibitem{BDFFKS} D.~Becirevic, I.~Dorsner, A.~Faroughy, S.~Fajfer, N.~Kosnik and O.~Sumensari, arXiv:1806.05689 [hep-ph]. 


\bibitem{Lattice:2015rga}
  J.~A.~Bailey {\it et al.} [MILC],
  Phys.\ Rev.\ D {\bf 92} (2015) no.3,  034506
  [arXiv:1503.07237 [hep-lat]];
H.~Na {\it et al.} [HPQCD],
  Phys.\ Rev.\ D {\bf 92} (2015) no.5,  054510
   Erratum: [Phys.\ Rev.\ D {\bf 93} (2016) no.11,  119906]
  [arXiv:1505.03925 [hep-lat]];
see also  M.~Atoui {\it et al.}, 
  Eur.\ Phys.\ J.\ C {\bf 74} (2014) no.5,  2861
  [arXiv:1310.5238 [hep-lat]].


\bibitem{deDivitiis:2008df}
  G.~M.~de Divitiis {\it et al.},
  Nucl.\ Phys.\ B {\bf 807} (2009) 373
  [arXiv:0807.2944 [hep-lat]].


\bibitem{CLN}
  I.~Caprini, L.~Lellouch and M.~Neubert,
  Nucl.\ Phys.\ B {\bf 530} (1998) 153
  [hep-ph/9712417];
C.~G.~Boyd, B.~Grinstein and R.~F.~Lebed,
  Phys.\ Rev.\ D {\bf 56} (1997) 6895
  [hep-ph/9705252].

\bibitem{EXP}
  A.~Abdesselam {\it et al.} [Belle],
  arXiv:1702.01521 [hep-ex];
B.~Aubert {\it et al.} [BaBar],
  Phys.\ Rev.\ D {\bf 77} (2008) 032002
  [arXiv:0705.4008 [hep-ex]].


\bibitem{Bigi:2017jbd}
  D.~Bigi, P.~Gambino and S.~Schacht,
  JHEP {\bf 1711} (2017) 061
  [arXiv:1707.09509 [hep-ph]];
B.~Grinstein and A.~Kobach,
  Phys.\ Lett.\ B {\bf 771} (2017) 359
  [arXiv:1703.08170 [hep-ph]].


\bibitem{Bouchard:2013pna}
  C.~Bouchard {\it et al.} [HPQCD],
  Phys.\ Rev.\ D {\bf 88} (2013) no.5,  054509
   Erratum: [Phys.\ Rev.\ D {\bf 88} (2013) no.7,  079901]
  [arXiv:1306.2384 [hep-lat]];
J.~A.~Bailey {\it et al.},
  Phys.\ Rev.\ D {\bf 93} (2016) no.2,  025026
  [arXiv:1509.06235 [hep-lat]].

\bibitem{Bordone:2016gaq}
  M.~Bordone, G.~Isidori and A.~Pattori,
  Eur.\ Phys.\ J.\ C {\bf 76} (2016) no.8,  440
  [arXiv:1605.07633 [hep-ph]].



\bibitem{Becirevic:2009fy}
  D.~Becirevic and N.~Kosnik,
  Acta Phys.\ Polon.\ Supp.\  {\bf 3} (2010) 207
  [arXiv:0910.5031 [hep-ph]]; 
S.~de Boer, T.~Kitahara and I.~Nisandzic,
  arXiv:1803.05881 [hep-ph].


\bibitem{Amhis:2016xyh}
  Y.~Amhis {\it et al.} [HFLAV],
  Eur.\ Phys.\ J.\ C {\bf 77} (2017) no.12,  895
  [arXiv:1612.07233 [hep-ex]], see HFLAV website for the most recent updates.


\bibitem{Aaij:2017tyk}
  R.~Aaij {\it et al.} [LHCb],
  Phys.\ Rev.\ Lett.\  {\bf 120} (2018) no.12,  121801
  [arXiv:1711.05623 [hep-ex]].



\bibitem{Aaij:2014ora}
  R.~Aaij {\it et al.} [LHCb],
  Phys.\ Rev.\ Lett.\  {\bf 113} (2014) 151601
  [arXiv:1406.6482 [hep-ex]].

\bibitem{Aaij:2017vbb}
  R.~Aaij {\it et al.} [LHCb],
  JHEP {\bf 1708} (2017) 055
  [arXiv:1705.05802 [hep-ex]].





\bibitem{Fajfer:2012vx}
  S.~Fajfer, J.~F.~Kamenik and I.~Nisandzic,
  Phys.\ Rev.\ D {\bf 85} (2012) 094025
  [arXiv:1203.2654 [hep-ph]].



\bibitem{us}
D.~Becirevic, D.~Leljak, B.~Melic and O.~Sumensari, {\it in preparation}; 
B.~Colquhoun {\it et al.} [HPQCD],
  PoS LATTICE {\bf 2016} (2016) 281
  [arXiv:1611.01987 [hep-lat]].
  
  

\bibitem{langenbruch} C.~Langenbruch, ``{\it Lepton flavour universality (LFU) tests in B decays as a probe for new physics}", talk presented at this conference.



\bibitem{LFUV-moriond}  A.~Greljo, ``{\it B-anomalies: A Model Builder's Guide }"; R.~Alonso, ``{\it Gauged flavour symmetry for B anomalies}; 
J.~Fuentes-Martin, ``{\it UV-complete model for B anomalies and SM flavor hierarchies}"; M.~Nardecchia, ``{\it Flavour anomalies and UV completion after $R_{K^{ (\ast)}}$}", talks presented at this conference.


\bibitem{Descotes-Genon:2015uva}
  S.~Descotes-Genon, L.~Hofer, J.~Matias and J.~Virto,
  JHEP {\bf 1606} (2016) 092
  [arXiv:1510.04239 [hep-ph]].


\bibitem{Ball:2004rg}
  P.~Ball and R.~Zwicky,
  Phys.\ Rev.\ D {\bf 71} (2005) 014029
  [hep-ph/0412079].



\bibitem{Gonzalez-Alonso:2017iyc}
  M.~Gonz\'alez-Alonso {\it et al.},
  Phys.\ Lett.\ B {\bf 772} (2017) 777
  [arXiv:1706.00410 [hep-ph]].


\bibitem{Wise:2014oea}
  M.~B.~Wise and Y.~Zhang,
  Phys.\ Rev.\ D {\bf 90} (2014) no.5,  053005
  [arXiv:1404.4663 [hep-ph]].


\bibitem{Faroughy:2016osc}
See e.g.  D.~A.~Faroughy, A.~Greljo and J.~F.~Kamenik,
  Phys.\ Lett.\ B {\bf 764} (2017) 126
  [arXiv:1609.07138 [hep-ph]].


\bibitem{Grygier:2017tzo}
  J.~Grygier {\it et al.} [Belle Collaboration],
  Phys.\ Rev.\ D {\bf 96} (2017) no.9,  091101
   Addendum: [Phys.\ Rev.\ D {\bf 97} (2018) no.9,  099902]
  [arXiv:1702.03224 [hep-ex]].




\bibitem{Becirevic:2016zri}
  D.~Be\v{c}irevi\'c, O.~Sumensari and R.~Zukanovich Funchal,
  Eur.\ Phys.\ J.\ C {\bf 76} (2016) no.3,  134
  [arXiv:1602.00881 [hep-ph]].


\bibitem{Alonso:2016oyd}
  R.~Alonso, B.~Grinstein and J.~Martin Camalich,
  Phys.\ Rev.\ Lett.\  {\bf 118} (2017) no.8,  081802
  [arXiv:1611.06676 [hep-ph]].


\bibitem{Dorsner:2017ufx}
  I.~Dor\v{s}ner {\it et al.},
  JHEP {\bf 1710} (2017) 188
  [arXiv:1706.07779 [hep-ph]].


\end{thebibliography}
\end{document}